\documentclass[10pt]{article}

\usepackage[
    a4paper,
    top=1.8cm,
    bottom=2.0cm,
    left=1.7cm,
    right=1.7cm
]{geometry}

\usepackage{microtype}

\usepackage{amsmath,amssymb}

\usepackage{graphicx}
\usepackage[font=small,labelfont=bf,labelsep=period]{caption}

\usepackage[numbers,sort&compress,super]{natbib}

\usepackage[hidelinks]{hyperref}

\usepackage[dvipsnames]{xcolor}

\title{\bfseries Wearable metasurfaces for boosting the effective 
area 
of mobile-device antennas  
}

\author{%
\begin{minipage}{0.95\textwidth}
\centering
\small
Mohammad M. Asgari$^{1,\dagger}$,
Xianghui Qiu$^{2,\dagger}$,
Francisco S. Cuesta$^{1}$,
Shuai S. A. Yuan$^{1}$,\\
Kamil A. I\c{s}\i k$^{1}$,
Sergei Tretyakov$^{1}$,
Xuchen Wang$^{2,*}$,
Viktar Asadchy$^{1}$
\end{minipage}%
}

\date{}

\begin{document}

\twocolumn[
\begin{@twocolumnfalse}

\maketitle


\noindent\textbf{Affiliations}\\
$^{1}$Department of Electronics and Nanoengineering, Aalto University, Espoo, Finland.\\
$^{2}$College of Physics and Optoelectronic Engineering, Harbin Engineering University, Harbin, China.\par\vspace{0.5em}

$^{\dagger}$These authors contributed equally

\textbf{Corresponding author:}\\
$^{*}$Xuchen Wang, xuchen.wang@hrbeu.edu.cn

\begin{abstract}
Wireless communications increasingly face scenarios with compact user equipments operating in propagation environments where signal blockage, absorption, and device size limitations strongly constrain link performance. Most approaches to wireless-link enhancement focus either on improving base-station antenna systems, for example, through massive MIMO architectures; on engineering the propagation environment using reconfigurable intelligent surfaces and active relays. Here, we propose an alternative user-side strategy for improving wireless links based on wearable metasurfaces that enhance the effective antenna area of compact wireless devices. The proposed passive metasurface is integrated into clothing and engineered to collect the power of electromagnetic waves incident on user’s body and route the collected power toward a mobile device in the form of surface waves. This mechanism significantly increases the effective receiving and transmitting area of the device antenna and its gain without external power sources, active control electronics, or additional stand-alone hardware. We implement the concept on a hoodie textile using a scalable screen-printing process with conductive silver ink and demonstrate an approximately tenfold enhancement of the received signal at 26~GHz. This passive approach effectively brings the human body into the communication network by turning clothing into a virtual  thin and flexible lens for focusing the incident power on the user equipment.
\end{abstract}

\vspace{1em}

\end{@twocolumnfalse}
]

\section*{Introduction}

\begin{figure}[tb]
  \centering
  \includegraphics[width=0.98\linewidth]{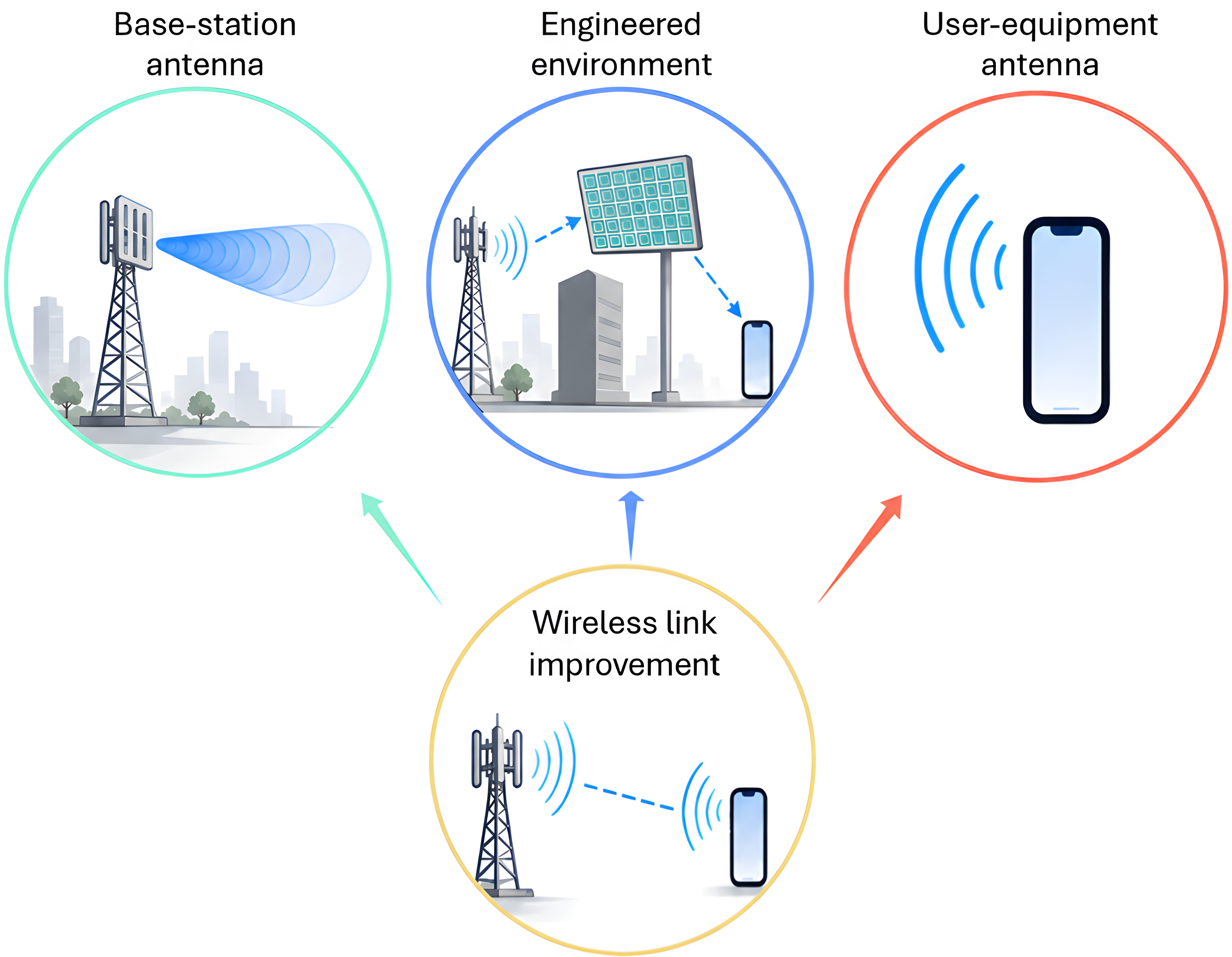}
  \caption{\textbf{Three complementary strategies for improving wireless links.} The link budget can be enhanced by increasing the gain of the base-station-side antenna system, by engineering the propagation environment, or by improving the gain or effective area of the user-equipment-side antenna system.}
  \label{fig:fig1}
\end{figure}

The growing demand for wireless data transmission and larger bandwidths is pushing next-generation communication systems toward millimeter-wave and sub-terahertz frequency bands. However, at these frequencies, the use of radio-wave propagation becomes increasingly challenging: free-space path loss increases with the frequency, while blockage, absorption, and scattering by objects in the environment become more severe~\cite{6732923,8732419,6515173}. In line-of-sight links, the Friis transmission formula shows that the received power is proportional to the gains of both the transmitting and receiving antennas~\cite{friis1946note}. In non-line-of-sight scenarios, an additional degree of freedom becomes available: the propagation environment itself can be engineered to create controlled reflection, redirection, or focusing paths that enhance the received signal~\cite{basar2019wireless,direnzo2020smart,kosulnikov2023simple}. From this perspective, wireless-link enhancement can be broadly classified into three complementary but fundamentally different strategies illustrated in Fig.~\ref{fig:fig1}: improving the base-station-side antenna system,  engineering the wireless environment between the two terminals, and improving the user-equipment-side antenna system.

The first strategy, aiming to improve the base-station-side antenna system, has been extensively pursued through electrically large aperture and leaky-wave antennas, phased arrays, beamforming architectures, and massive-MIMO systems~\cite{heath2016overview,Jackson2012LeakyWaveAntennas,oliner2007leaky}.  
The second strategy, based on engineering or assisting the propagation environment, has been actively explored using reconfigurable intelligent surfaces (RISs), programmable metasurfaces, relays, and smart repeaters, which enhance wireless links by redirecting, reshaping, or re-radiating electromagnetic waves along more favorable propagation paths~\cite{huang2019reconfigurable,bjornson2019intelligent,di2020smart,carter2024flatknit,asgari2024multifunctional,zhu2025selfcontrolled,yang2025adaptively}. 
Compared with these approaches, the third strategy, targeting on enhancing the user-equipment-side antenna system capabilities, has received much less attention as a route for link-budget enhancement. 

Considerable effort has been devoted to beamforming and MIMO antenna configurations for mobile terminals, especially at millimeter-wave frequencies~\cite{hong2017millimeter,zhang2017fiveg}. However, user equipment is subject to much stricter constraints on physical size, form factor, power consumption, hardware complexity, and interaction with the human body. In particular, phones and wearable devices provide only a limited area for antenna integration, which restricts the achievable effective transmitting/receiving area of the antenna and its radiated power~\cite{li2018mmwave,harrington1959effect,8012469}.
Another approach is to place high-impedance surfaces or artificial magnetic conductors near the user-equipment antenna to suppress backward radiation into the body and enhance the realized gain in on-body scenarios~\cite{yan2014low,ashyap2017compact}. However, such structures do not overcome the limited-aperture constraint of compact user equipment and require careful co-design with the antenna and body-worn configuration.

In this paper, we propose an alternative route to enhance the performance of user-equipment antennas without external power sources, active electronic control, or additional stand-alone devices carried by the user. The proposed concept is based on a wearable textile metasurface with an ink-printed pattern. 
This passive wearable metasurface is engineered to perform two functions simultaneously: it converts electromagnetic waves incident on the user’s body into guided surface waves, and it routes these waves toward a designated region of the garment, such as a pocket or sleeve, where a mobile device may be located during idle or active use. As a result, the mobile antenna, whose effective transmitting/receiving area is normally limited by the compact size of the user equipment, can receive not only the direct signal from the base station but also an additional guided signal contribution collected over a much larger area of the textile-covered body. This additional signal would otherwise be mostly absorbed by biological tissue or scattered away from the receiver. In this way, the wearable metasurface increases the effective area and gain of a nearby user-equipment antenna in both reception and transmission regimes, owing to the reciprocity of the passive metasurface system.
As a specific implementation, we design and fabricate a metasurface on a textile using a low-cost and scalable screen-printing process with conductive silver ink. The printed metasurface concentrates incident waves along a designated extended region, where a compact subwavelength antenna operating at 26~GHz is placed. Through novel theoretical modeling, metasurface design, fabrication, and experimental characterization, we demonstrate an approximately tenfold enhancement of the effective area of the antenna. 
Therefore, the proposed metasurface-based clothing effectively transforms the human body into a part of the communication network, concentrating the incident  power on the user equipment.

\section*{Results}

\subsection*{Concept}

At millimeter-wave frequencies, the user side of a wireless link is often limited not only by propagation loss and blockage, but also by the small 
size of 
compact mobile devices. The importance of this  limitation can be understood from the Friis transmission formula~\cite{friis1946note}. For a far-field link under polarization- and impedance-matched conditions, the received power at the mobile device can be written as $P_{\rm rx}=P_{\rm tx}G_{\rm tx}A_{\rm e,rx}/(4\pi R^2)$, where $P_{\rm tx}$ and $P_{\rm rx}$ are the transmitted and received powers, $G_{\rm tx}$ is the transmitting (base station) antenna gain in the link direction, $R$ is the link distance, and $A_{\rm e,rx}$ is the effective receiving area of the user-equipment antenna. Thus, for a fixed propagation distance and a given base-station-side antenna, the received power scales linearly with the effective antenna area available at the user side.

\begin{figure*}[t]
  \centering
  \includegraphics[width=\linewidth]{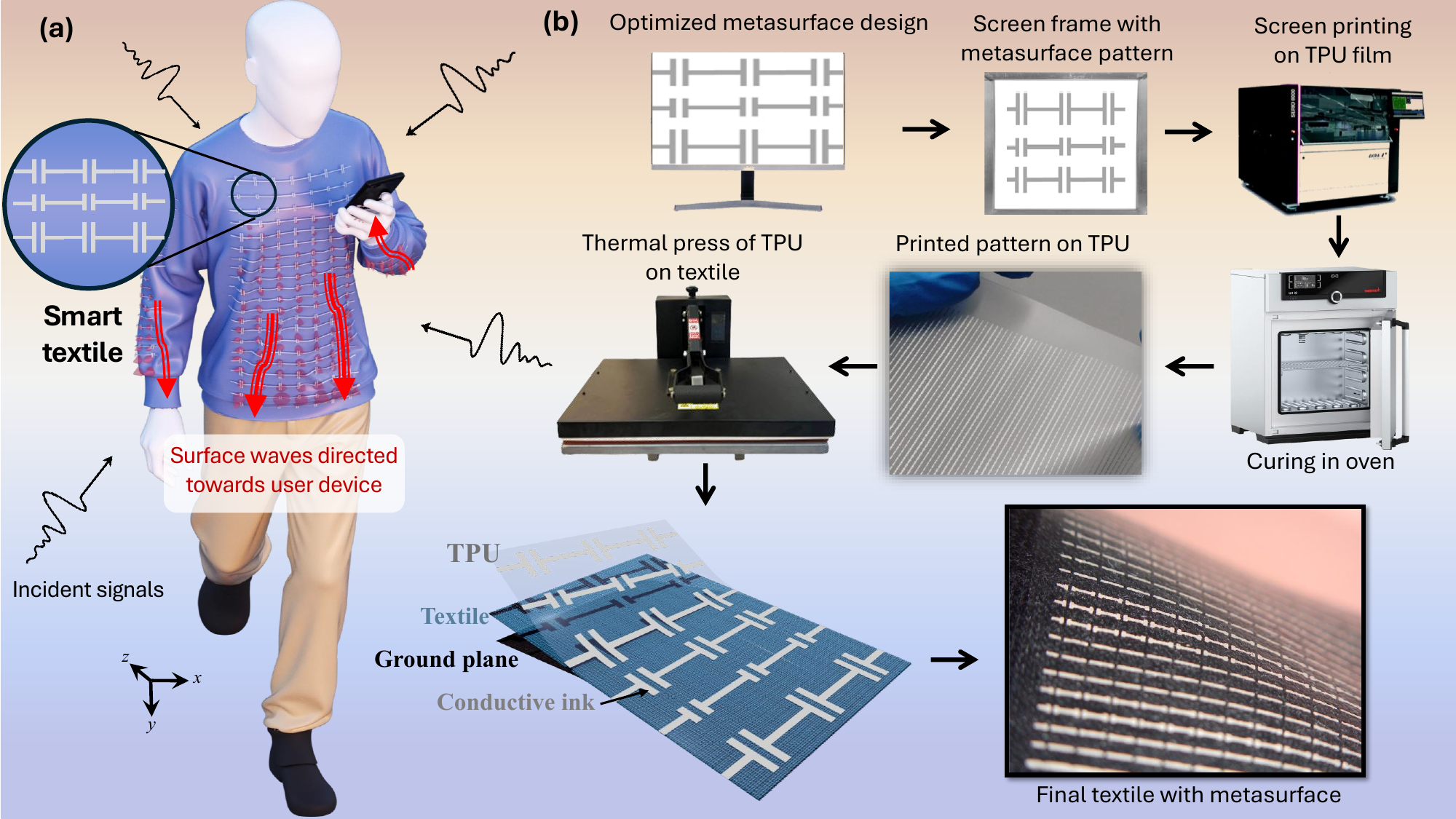}
  \caption{\textbf{Concept and textile fabrication workflow.} \textbf{a,} Smart garment with an integrated metasurface that intercepts incident radio-frequency waves on the torso, converts them into guided surface waves and redirects energy toward pockets and/or cuffs where compact antennas are typically located. \textbf{b,} Scalable manufacturing route: the metasurface pattern is designed using a semi-analytical approach, transferred to a screen (mask), printed with  conductive (silver) ink onto a TPU carrier film, thermally cured, and finally heat-pressed onto the fabric to yield a conformal, wash-durable conductive pattern.}
  \label{fig:fig2}
\end{figure*}

In millimeter-wave mobile antennas, this area is severely constrained by the physical size and form factor of the device. Typical 5G smartphone antenna modules operating near the 26--28 GHz band have the effective receiving areas on the order of $1~{\rm cm}^2$ (see e.g., Refs.~\cite{Zhang2017PlanarSwitchable3DCoverage,Zhang2018CompactBeamSteerable}). This small area directly limits the power that can be collected from the incident waves. By reciprocity, the same antenna has the same effective area when operating in transmission, meaning that compact user-equipment antennas are similarly constrained in their ability to radiate power efficiently toward the base station. Consequently, both the downlink and uplink suffer from the same fundamental antenna bottleneck at the user side.

Our approach is to overcome this bottleneck by using the much larger surface area that is already present around the user. The front and back sides of the torso, together with the sleeves, can provide a textile-covered area approaching the square-metre scale, which is several orders of magnitude larger than the effective area of a compact millimeter-wave antenna. We propose to embed judiciously designed passive metasurfaces into clothing so that electromagnetic waves incident on the body are not merely absorbed by biological tissue or scattered away, but are instead collected over the garment area, converted into guided surface waves, and routed toward a nearby mobile device. In reception, the antenna therefore collects not only the direct incident signal but also an additional contribution gathered over the extended textile surface, guided along the garment, and delivered to the vicinity of the device by the metasurface. 

At the level of electromagnetic functionality, this concept builds on the broader metasurface paradigm of converting propagating waves into bound or guided surface waves. Such plane-wave-to-surface-wave conversion has been extensively explored in rigid microwave metasurfaces, typically implemented as stand-alone planar structures on printed-circuit-board \cite{sun2012gradient,wan2014simultaneous,kwon_arbitrary_2018,achouri2018space,Popov2019omega,kwon_modulated_2020,xu_wideangle_2022,xu_extreme_2022,arshed2024direct,budhu2024near,Arshed2025SurfaceWaveMetaprism,Li2026SuperdirectiveSpaceToSurface}. However, the achievable efficiency of such designs was often limited by finite-aperture scattering, parasitic radiation channels, and impedance-discretization constraints. Here, we translate this wave-conversion principle into a different physical and system-level setting: a passive, screen-printed textile platform that uses the garment itself as an effective-area-expansion layer for compact user equipment. 

To design such a platform, we develop a new Green-function-based optimization framework that explicitly accounts for the finite size of the metasurface, the presence of a grounded lossy dielectric substrate, and absorption in the finite-conductivity printed ink pattern. This is essential for textile implementations, where the useful guided power must be maximized not in an ideal infinite structure, but in a finite, lossy, and manufacturable device. 
We introduce several figures of merit (FOM) for quantifying passive signal enhancement by smart textiles and examine how optimizing the metasurface with respect to different metrics affects the final balance between field enhancement, surface-wave capture, absorption, parasitic specular reflection, and edge scattering. These metrics are defined with the practical device location in mind: the useful power is not simply the power coupled into a surface wave, but the portion of that power delivered to a region where a compact antenna can actually be placed.
The mobile device may be a smartphone carried in a pocket or held in the hand, a portable radio, a wearable sensor, a headset, or another compact wireless terminal located close to the textile. 
This operating principle is illustrated in Fig.~\ref{fig:fig2}(a).

The metasurface clothing acts as a passive antenna area-expansion layer for user equipment. Even if only a fraction of the available garment area is efficiently coupled to the antenna, the increase in usable area can reach one or several orders of magnitude, offering a route to substantially improved received power in coverage-limited environments. The same structure also operates reciprocally: when the mobile device transmits, the guided fields excited on the metasurface are radiated from the extended textile area, improving the effective uplink radiation toward the surrounding wireless infrastructure.

From the antenna-system perspective, the combination of the textile metasurface and the nearby compact user-equipment antenna forms an effective leaky-wave area. However, this architecture differs fundamentally from conventional stand-alone leaky-wave antennas \cite{Jackson2012LeakyWaveAntennas,Xu2013UnderstandingLeakyWaveStructures,Javanbakht2021ReviewReconfigurableLWA}. Conventional leaky-wave antennas require a dedicated feed, a purposely built radiating structure, surface-wave transport infrastructure, and additional device-level integration~\cite{Jackson2012LeakyWaveAntennas,Lee2024PrintedLWA,Budhu2024ConformalSurfaceWave,Arshed2025SurfaceWaveMetaprism}.
Here, by contrast, the feed is the antenna that already exists in the handheld or wearable device, without being mechanically attached to the metasurface. Because of that, the whole textile surface is optimized to concurrently maximize radiating-to-surface wave conversion and surface-wave launching and guiding towards the antenna of a handheld device, avoiding splitting the surface into receiving, transporting, and transmitting subsections. 
The metasurface therefore transforms a compact mobile antenna into a large-effective-area system without requiring a new active or passive antenna module, a wired connection to the device, or additional user-carried hardware.

The proposed concept is also fundamentally distinct from active wearable antenna systems \cite{Locher2006TextilePatchAntennas,Salvado2012TextileMaterialsWearableAntennas,Paracha2019WearableAntennasReview}. The metasurface textile has no wired connection to the user equipment. It is completely passive: it does not contain amplifiers, phase shifters, batteries, control circuits, or communication electronics. Its function is implemented entirely through the spatially engineered electromagnetic response of the printed conductive pattern. 
To benefit from the metasurface cloth, the user equipment does not need any modifications and can be used as usual.

In this work, as a proof-of-concept, we intentionally consider a simplified but practically relevant configuration. The metasurface is designed to operate at $26~{\rm GHz}$, within the millimeter-wave frequency range, for normally incident waves with a fixed linear polarization. Specifically, we consider transverse-electric (TE) polarization with respect to the plane defined by the incident propagation direction and the desired surface-wave propagation direction. We further restrict the metasurface to a one-dimensional design that is uniform along the horizontal direction in Fig.~\ref{fig:fig2}(a) and homogeneously guides the collected power toward the lower edge of the garment. This choice creates an extended enhancement region near the bottom of clothing, where smartphones, hand-held radios, and other portable devices are often located in practice, for example, in trouser pockets. These simplifications allow the aperture-expansion mechanism to be demonstrated clearly, without obscuring the physical principle by the additional complexity of a fully conformal, multi-angle, and polarization-diverse garment.

Having established this principle of effective area expansion, 
the remaining challenge is to realize a large-area 
functional 
surface that is scalable, flexible, sufficiently low-loss, and compatible with everyday textiles. We identify screen printing as a particularly suitable fabrication route for this purpose \cite{Komolafe2021ETextileReview}. Screen printing is a mature industrial process that is routinely used to pattern large-area textiles, including shirts, jackets, hoodies, and other garments, with graphic designs and functional coatings. 
Compared with yarn-level approaches such as flat-knit textile metasurfaces \cite{carter2024flatknit}, screen printing decouples the electromagnetic pattern from the textile microstructure: the conductive layout can be designed with impressive 170-micron geometric resolution and then transferred onto a wide range of commercially available fabrics without being constrained by stitch size, yarn routing, float irregularity, or knitting-machine architecture. This fabrication flexibility also enables practical deployment scenarios beyond direct integration into everyday garments. For example, the metasurface could be printed on a separate lightweight vest or textile overlay that can be worn over existing clothing, rather than being fabricated individually on each garment. Such a modular implementation would further reduce manufacturing cost, simplify cleaning and replacement, and make the technology more convenient for users.

In the present implementation, the conventional gra\-phic ink used in textile printing is replaced by a conductive silver-flake ink (Metalon  HPS U57B by NovaCentrix), allowing the printed pattern to function as the metallic layer of the metasurface. This approach combines the scalability and garment compatibility of industrial textile printing with the geometric precision required for millimeter-wave metasurface design. 
%
%
Our screen-printing fabrication workflow is illustrated in Fig.~\ref{fig:fig2}(b). Starting from the computationally optimized metasurface design, the corresponding metallic layout is transferred to a screen-printing frame, which defines the pattern during printing. A conductive silver-flake ink is then deposited onto a $100~\mu$m-thick thermoplastic po\-ly\-ure\-tha\-ne (TPU) carrier film (LPT5302 MR7X by Covestro), chosen both as a printable support and as a transfer layer for subsequent textile integration. After thermal curing, which removes residual solvent and improves the conductivity and mechanical stability of the printed traces, the patterned TPU film is aligned with the fabric and bonded by heat pressing. More information on the metasurface textile fabrication can be found in Methods. In the final structure, the printed conductive pattern, the TPU film, textile substrate, and ground-plane layer form a flexible grounded metasurface that preserves the designed electromagnetic geometry while remaining compatible with garment-scale fabrication. 

For simplicity in the present prototype, the ground plane is implemented using a thin metallic backing layer. In practical garment implementations, however, this layer could also be realized by printing or laminating a second conductive TPU film on the opposite side of the textile, with a simple two-dimensional metallic mesh whose subwavelength wire spacing makes it behave effectively as a continuous ground plane at the operating frequency~\cite{tretyakov2003}. A close-up photograph of the fabricated hoodie-integrated metasurface is shown in the lower-right panel of Fig.~\ref{fig:fig2}(b), revealing high-quality print with resolution sufficient for creating metasurfaces operating at 26~GHz and above.


It is important to verify that smart textiles fabricated by screen printing can retain their physical integrity under demanding usability conditions associated with everyday wear. In particular, the conductive metasurface pattern must remain intact after textile laundering. We therefore performed multiple washability tests of the smart textile and observed high tolerance of the printed pattern even after multiple washing cycles, as detailed in Supplementary Section~1.

\begin{figure*}[t]
  \centering \includegraphics[width=\linewidth]{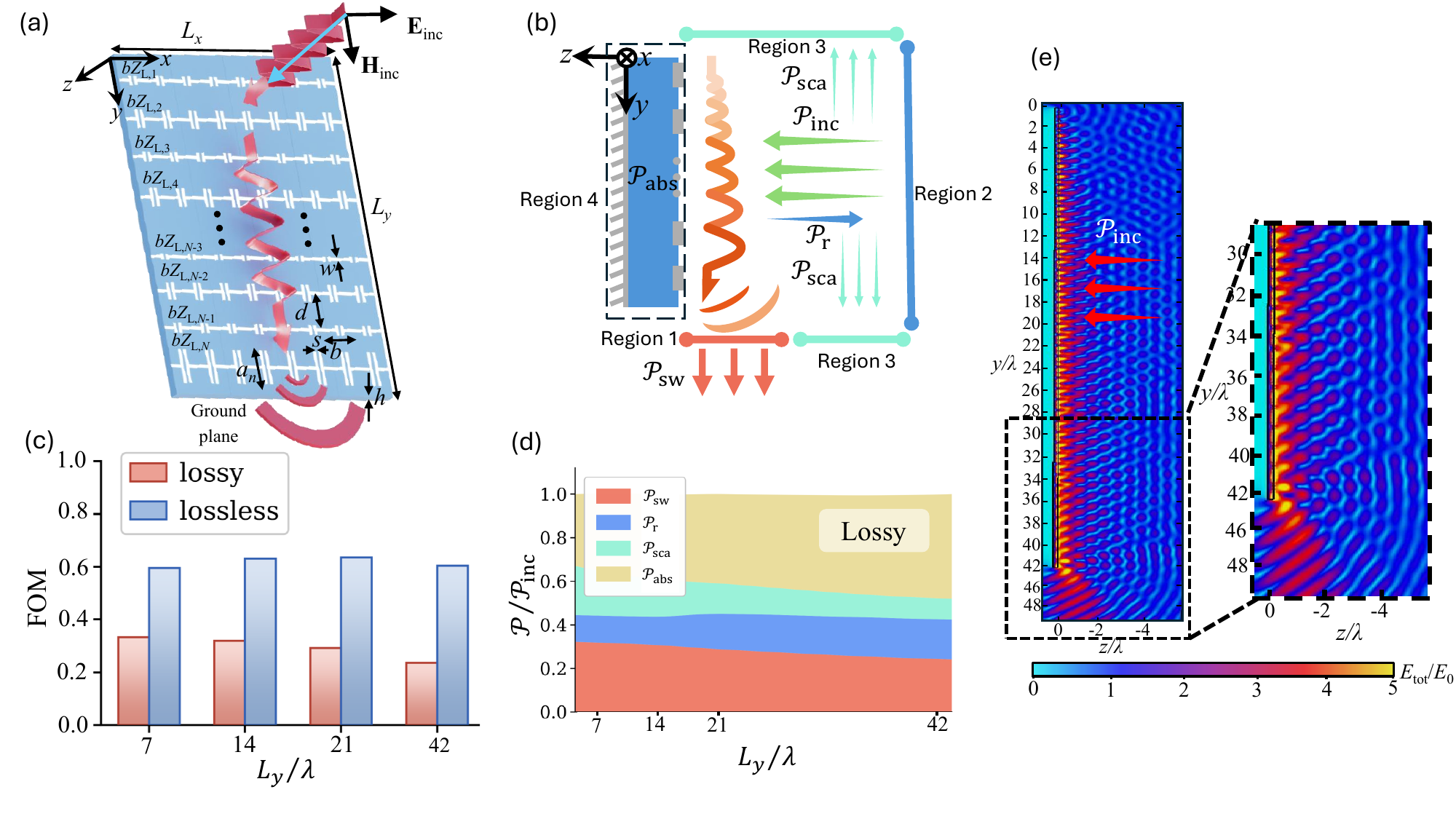}
  \caption{\textbf{Design and numerical performance of the aperiodic textile metasurface.} 
  \textbf{a}, Geometry of the metasurface formed by a conductive strip array on a grounded textile substrate. The structure is periodic along the $x$-direction with period $b$ and aperiodic along the $y$-direction, with each strip loaded by a capacitive unit cell parameterized by $a_n$. The incident wave is TE-polarized with the electric field oriented along the strips.
\textbf{b}, Schematic power-balance representation used to quantify the metasurface performance. The incident power $\mathcal{P}_{\rm inc}$ is redistributed into useful surface-wave power $\mathcal{P}_{\rm sw}$ directed toward the receiving region, specularly reflected power $\mathcal{P}_{\rm r}$, parasitically scattered power $\mathcal{P}_{\rm sca}$, and absorbed power $\mathcal{P}_{\rm abs}$.
\textbf{c}, Optimized figure of merit as a function of metasurface length $L_y$ for lossless and lossy implementations.
\textbf{d}, Power redistribution among $\mathcal{P}_{\rm sw}$, $\mathcal{P}_{\rm r}$, $\mathcal{P}_{\rm sca}$ and $\mathcal{P}_{\rm abs}$ for lossy metasurfaces with different lengths.
\textbf{e}, Simulated total electric-field amplitude for the realistic metasurface pattern with $L_y=42\lambda$, showing efficient conversion of the normally incident plane wave into a guided surface wave directed toward the lower edge of the textile area.}
  \label{fig:fig3}
\end{figure*}

\subsection*{Design and Figure of Merit}


We consider the metasurface structure shown in Fig.~\ref{fig:fig3}(a). The metasurface is uniform and periodic along the $x$-direction with subwavelength periodicity $b$, while it is aperiodic along the $y$-direction. The pattern consists of a conductive strip array with spacing $d$ along the $y$-direction located on a grounded substrate of thickness $h$ and relative permittivity $\varepsilon_{\rm s}$. Each strip of width $w$ is periodically loaded along the $x$-axis by reactive (capacitive) loads with the impedance density $Z_{\mathrm{L},n}$ (load impedance per unit length along the $x$-direction). As is shown below, such capacitive impedance can be realized for TE-polarized waves by splitting the strip into capacitor-type gaps with arm length $a_n$ [see Fig.~\ref{fig:fig3}(a)]. The width of the capacitive arms is $w$. The overall dimensions of the metasurface along the $x$ and $y$ directions are $L_x= M b $ and $L_y= N d$, respectively, featuring $N$ strips with $M$ capacitive loads in each strip.
The metasurface is assumed to be illuminated by a TE-polarized plane wave at an incident angle $\theta_{\rm i}$ with an electric field given by 
$\mathbf{E}_{\mathrm{inc}}(y,z)= E_0 e^{-j k_0(\sin\theta_{\rm i} y + \cos\theta_{\rm i} z)} \hat{\textbf{x}}$,
where $k_0$ is the wave number in free space and $E_0$ is the amplitude of the incident field. 
In what follows, without loss of generality, the frequency of the incident wave is chosen as $f = 26$~GHz, and we assume normal incidence ($\theta_{\rm i}=0$). 
Throughout the
paper, the time dependence of the complex-valued fields is assumed to be in the form of $e^{j \omega t}$.

The conductive-pattern design is based on a semi-analytical model that combines a precomputed numerical Green function with a circuit model of the loaded strips \cite{PhysRevApplied.14.044007}. After the background structure has been characterized once by full-wave simulations, the scattered fields produced by the metasurface with any given set of load impedances can be obtained rapidly by merely solving a finite system of linear equations. This avoids repeated full-wave simulations during the optimization loop and makes it possible to optimize large finite metasurfaces with many independently loaded strips. Importantly, the model accounts for  the finite size of the metasurface and the grounded dielectric substrate as well as dissipation in both the substrate and the conductive pattern. Once the optimal load-impedance distribution is found, the corresponding printable unit-cell geometry is obtained through a mapping procedure based on the locally periodic approximation.

We model each conductive strip as a uniform line current directed along the $x$-axis. This approximation is valid when the strip width $w$ is much smaller than both the free-space wavelength and the inter-strip spacing, that is, $k_0 w \ll 1$ and $w \ll d$. Following the thin-wire model~\cite{tretyakov2003}, a narrow strip of width $w$ is represented by an equivalent cylindrical wire with the effective radius $r_{\rm eff}=w/4$. The line current on the $n$-th strip is denoted by $I_n$, and the strip center is located at $y_n=(n-1)d$ and $z_n=-h$, where the origin of the $z$ coordinate is chosen at the ground-plane level. The electric field radiated by the $n$-th line current at an observation point $(y,z)$ is written as $G_n(y,z)I_n$, where $G_n(y,z)$ is the numerically evaluated Green function of the grounded finite substrate for a source located at the $n$-th strip.

For each loaded strip, the induced current must satisfy Ohm's law. At the center of the $n$-th strip, this condition can be written as
\begin{equation}
Z_{{\rm L},n} I_n
=
E_{x,n}^{\rm ext}
+
E_{x,n}^{\rm self}
+
E_{x,n}^{\rm mut},
\label{eq:matrix_Ohm}
\end{equation}
where $E_{x,n}^{\rm ext}$ is the external field acting on the strip in the absence of the loaded strip array, while $E_{x,n}^{\rm self}$ and $E_{x,n}^{\rm mut}$ are the fields produced by the $n$-th strip itself and by all other strips, respectively. The external field includes the incident wave and the field reflected and diffracted by the finite grounded substrate, and is evaluated numerically using full-wave simulations of the background structure without the strips.

As is shown in Supplementary Section~2, the mutual and self-interaction field terms can be collected into an impedance matrix, which transforms Eq.~(\ref{eq:matrix_Ohm}) into a compact form:
\begin{equation}
\overline{\overline{\mathbf{Z}}}\cdot \mathbf{I}
=
\mathbf{E}^{\rm ext}_x,
\label{eq:impedance_matrix}
\end{equation}
where $\mathbf{I}=[I_1,I_2,\ldots,I_N]^{T}$ is the vector of strip currents and $\mathbf{E}^{\rm ext}_x=[E_{x,1}^{\rm ext},E_{x,2}^{\rm ext},\ldots,E_{x,N}^{\rm ext}]^{ T}$ is the external electric field at the strip positions. The impedance matrix $\overline{\overline{\mathbf{Z}}}$ contains the prescribed load impedances $Z_{{\rm L},n}$, the self-impedances of the finite-width strips, and the mutual impedances between all pairs of strips. These impedance terms are extracted from the numerically computed Green function of the current line on the finite grounded substrate, so that the effects of the substrate, finite area, and material losses are included in the optimization procedure~\cite{PhysRevApplied.14.044007}. For any trial set of load impedances, Eq.~\eqref{eq:impedance_matrix} directly gives the induced current distribution by a single linear-system solution, after which the scattered and total fields at every point in space can be evaluated.

The load-impedance distribution required for the desired metasurface functionality is found by numerical optimization. Specifically, we use a multiple-population genetic algorithm to optimize the set of reactive load impedances $Z_{{\rm L},n}$. This algorithm maintains population diversity by evolving several subpopulations in parallel and periodically exchanging individuals between them, which helps avoid premature convergence to local optima.

Different figures of merit can be introduced to quantify passive signal enhancement near a smart textile, depending on the assumed location, orientation, and type of the user-equipment antenna. In the main text, we focus on a robust metric defined as the fraction of the incident power directed toward a finite collection region where a mobile antenna can be placed:
\begin{equation}
\mathrm{FOM}
=
\frac{1}{\mathcal{P}_{\rm inc}}
\int_{-h-h_1}^{-h}
S_y^{\rm tot}(y_{\rm s},z){\rm d}z .
\label{eq:fomp}
\end{equation}
Here, $S_y^{\rm tot}=-\Re \{E_x^{\rm tot}H_z^{{\rm tot}*} \}/2$ is the $y$-component of the time-averaged Poynting vector for the total field, including the incident field and all scattered fields. The integration is performed over linear Region~1 indicated in Fig.~\ref{fig:fig3}(b), located at $y_{\rm s}=L_y+\lambda$ and extending from $z=-h-h_1$ to $z=-h$. Here, we assume the metasurface is periodic along the $x$-direction. We choose $h_1=\lambda$, corresponding to one free-space wavelength at 26~GHz. The normalization factor $\mathcal{P}_{\rm inc}$ is the incident power illuminating the metasurface per unit length along the $x$-direction. Additional details on the definition and numerical evaluation of Eq.~\eqref{eq:fomp} are provided in Supplementary Section~3.

This FOM is particularly useful when the exact position and orientation of the mobile antenna are not fixed, as is typical for practical user equipment placed near clothing. Maximizing the FOM defined in Eq.~\eqref{eq:fomp} increases the power delivered to an extended region rather than to a single point, thereby improving the average probability that a nearby compact antenna can benefit from the metasurface. In Supplementary Section~4, we also analyze three alternative FOM definitions and compare the corresponding optimized metasurface designs. These alternative metrics may be preferable when the antenna position, orientation or receiving mode is known in advance.

We next optimize the metasurface parameters to maximize the FOM in Eq.~\eqref{eq:fomp}. For an example  design, we choose the strip width to be $w=\lambda/43$, the fixed gap distance $s=\lambda/40$, and the center-to-center spacing between adjacent strips  $d=\lambda/6$. The period along the $x$-direction is set to $b=\lambda/5.5$, ensuring a 
homogenizable 
response along the direction in which the structure is periodic. The textile substrate is modeled as a cotton--polyester blended fabric with the relative permittivity $\varepsilon_{\rm s}=1.81$ and loss tangent $\tan\delta_{\rm s}=0.015$, estimated by interpolation from Ref.~\cite{Ouyang2008High} for a cotton--polyester ratio of 30--70. The lossy substrate is represented by the complex permittivity $\varepsilon_{\rm s}(1-j\tan\delta_{\rm s})$. The substrate thickness was measured to be approximately $h=2.2$~mm.

The bulk resistivity of the screen-printed silver ink was measured to be $\rho=0.0716~\Omega\cdot\mu{\rm m}$, as described in Methods. This conductor loss is represented by an equivalent series resistance per unit length added to each loaded strip, so that $Z_{{\rm L},n}
=
R_{\rm loss}
+
jX_{{\rm L},n}$,
where $X_{{\rm L},n}$ is the reactive load impedance density. The equivalent loss resistance is estimated as $R_{\rm loss}=8.5\times10^3~\Omega/{\rm m}$. Since the loads are capacitive, the optimized reactance values are negative and constrained to the physically realizable range $ -5\times10^5~\Omega/{\rm m}
<
X_{{\rm L},n}
<
-1.8\times10^5~\Omega/{\rm m}$.
This constraint ensures that the optimized impedance distribution can be implemented using the selected printed capacitive unit-cell geometry.
After the optimal load reactances $X_{{\rm L},n}$ are obtained, the loaded-strip model is converted into a printable metasurface pattern. Specifically, each optimized load impedance is mapped to a corresponding capacitive unit-cell geometry, parameterized by the arm length $a_n$, using a locally periodic approximation. The details of this impedance-to-geometry mapping procedure, as well as the optimized values of $X_{{\rm L},n}$ and $a_n$ are provided in Supplementary Section~5.

To verify the effectiveness of the design framework, we optimized textile metasurfaces with four different lengths: $L_y=7\lambda$, $14\lambda$, $21\lambda$ and $42\lambda$, corresponding to $N=42$, $84$, $126$ and $252$ loaded strips, respectively. Figure~\ref{fig:fig3}(c) shows the optimized FOM values defined by Eq.~\eqref{eq:fomp} for these four length sizes. The blue bars correspond to the ideal lossless case, whereas the red bars show the values that include losses in both the textile substrate and the printed conductive ink. In the lossless case, the FOM reaches values close to 0.6 and remains nearly constant as the metasurface length increases. This indicates that approximately 60\% of the power incident on the metasurface area can be redirected toward the receiving region near the lower edge of the textile.
When material losses are included, the FOM is reduced, as expected, and gradually decreases with increasing metasurface length. This trend occurs because a longer metasurface  excites  surface waves over a larger area, increasing the propagation length over the lossy textile and therefore the accumulated absorption. Nevertheless, the absolute power delivered toward the receiving edge still increases with $L_y$, because the incident power collected by the larger metasurface also increases. Thus, larger textile samples remain beneficial, although the gain is sublinear in the presence of absorption.

To clarify how the incident power is redistributed, we evaluate the power balance for the lossy metasurface designs, as summarized in Fig.~\ref{fig:fig3}(d). Note that all the power quantities represent power per unit length along the $x$-direction of the metasurface, and therefore, denoted as $\mathcal{P}$ rather than $P$. The incident power can be sent into four main channels: useful surface-wave power delivered toward the receiving region, $\mathcal{P}_{\rm sw}$; specularly reflected power, $\mathcal{P}_{\rm r}$; parasitic power scattered into undesired directions, $\mathcal{P}_{\rm sca}$; and absorbed power, $\mathcal{P}_{\rm abs}$. The quantities $\mathcal{P}_{\rm sw}$, $\mathcal{P}_{\rm r}$ and $\mathcal{P}_{\rm sca}$ are obtained by integrating the normal component of the total time-averaged Poynting vector over Region~1, Region~2, and Regions~3 in Fig.~\ref{fig:fig3}(b), respectively. The absorbed power $\mathcal{P}_{\rm abs}$ is evaluated from the net Poynting flux through the closed contour surrounding the metasurface and substrate, denoted as Region~4 in Fig.~\ref{fig:fig3}(b).

The power balance illustrated in Fig.~\ref{fig:fig3}(d) shows that absorption increases steadily with the metasurface length, confirming that material loss is the dominant limitation for large textile samples. At the same time, the relative contribution of edge scattering decreases for longer metasurfaces, because edge scattering becomes less important compared with 
the other effects. The specularly reflected power increases moderately with $L_y$, indicating residual imperfect cancellation of the reflected propagating wave over very long areas. This effect is not fundamental and can be reduced by increasing the optimization effort or using more advanced optimization methods. Overall, these results show that lower-loss textile substrates and higher-quality conductive inks can substantially improve the achievable FOM. However, even for the present, rather  lossy implementation, approximately one quarter to one third of the incident power can still be directed toward the receiving region.

Figure~\ref{fig:fig3}(e) shows the simulated total electric-field distribution for the practical  metasurface pattern with $L_y=42\lambda$ (the optimized values of gap widths $a_n$ can be found in Supplementary Section~5). The field map confirms that the designed textile metasurface converts the normally incident plane wave into a guided surface wave and channels the energy toward its lower edge. Near this receiving edge, the electric-field amplitude reaches values of approximately five times of the incident wave amplitude. Such  local field enhancement demonstrates the ability of the passive textile metasurfaces to substantially increase the signal available to a compact mobile antenna placed near the garment.


\begin{figure*}[t]
    \centering
    \includegraphics[width=\textwidth]{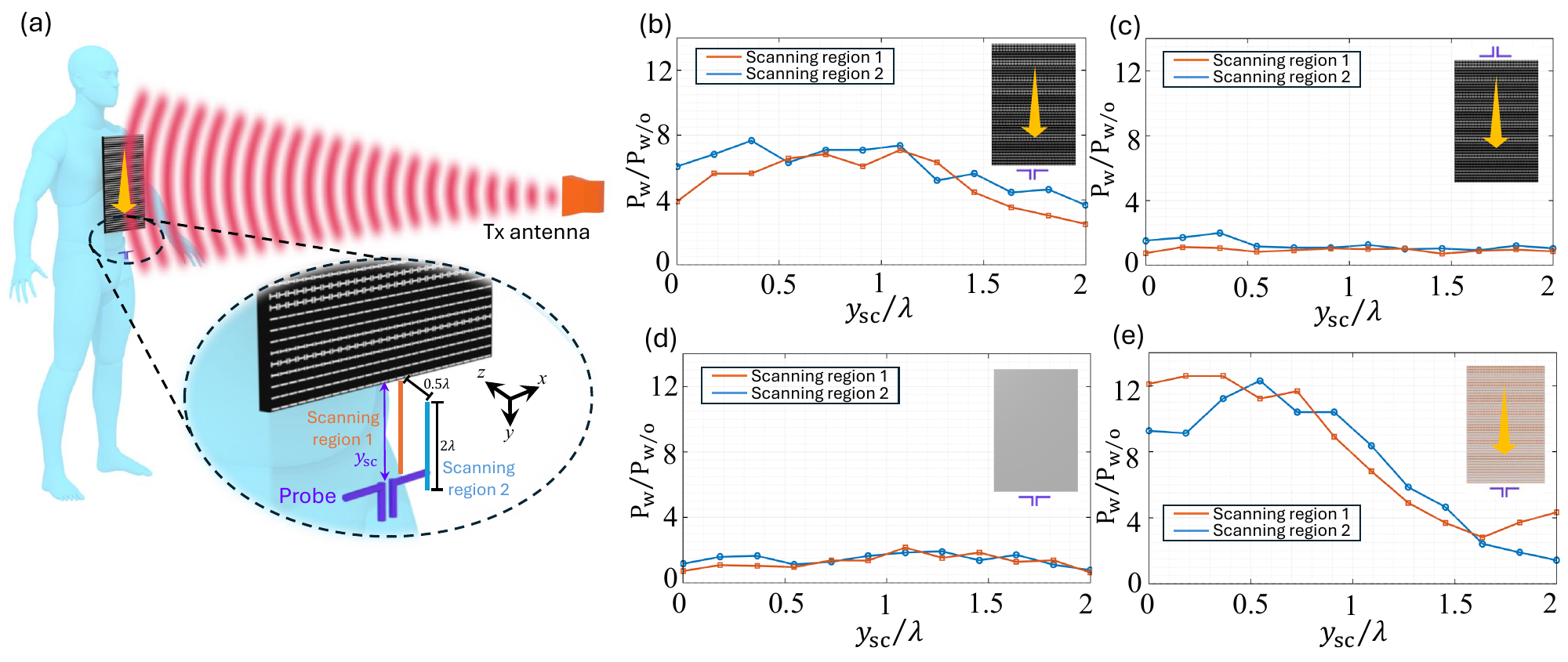}
    \caption{\textbf{Experimental validation of passive wireless-signal enhancement.}
\textbf{a,} Schematic of the measurement configuration. A transmitting antenna illuminates the metasurface sample, and a subwavelength probe measures the received power near the metasurface edge. The received power is recorded along two scan regions: scanning region~1 is located in the metasurface plane, whereas scanning region~2 is displaced from it by $0.5\lambda$ in the normal $-z$-direction. In both cases, the probe is scanned over a distance of $2\lambda$ from the metasurface edge, with coordinate $y_{\rm sc}$ measured along the scan direction. The mannequin is shown only to illustrate the envisioned body-worn configuration.
\textbf{b,} Measured received-power enhancement ratio $P_{\rm w}/P_{\rm w/o}$ near the intended receiving edge of the screen-printed textile metasurface. Here, $P_{\rm w}$ and $P_{\rm w/o}$ denote the received powers measured with and without the metasurface sample, respectively.
\textbf{c,} Control measurement performed near the opposite edge of the textile metasurface, confirming that the enhancement is directional and concentrated on the designed receiving side.
\textbf{d,} Control measurement for a copper plate with the same aperture size as the metasurface, confirming that the observed enhancement cannot be attributed merely to the presence of a conducting surface or edge scattering.
\textbf{e,} Corresponding measurement for a metasurface fabricated on a rigid printed circuit board, showing stronger peak enhancement under more controlled fabrication conditions. }
  \label{fig:fig4}
\end{figure*}

\subsection*{Experimental validation}

To validate the proposed concept experimentally, we measured the local signal enhancement produced by the fabricated metasurface near its lower edge, where a compact user-equipment antenna would be placed. 
The textile metasurface was fabricated with dimensions of $L_x=17.1\lambda=197$~mm and $L_y=40.4\lambda=466$~mm. 
The measurements were performed in a spacious university-building lobby rather than in an anechoic chamber, providing a realistic indoor propagation environment with background scattering from walls, furniture,  and surrounding objects. As depicted in Fig.~\ref{fig:fig4}(a), a transmitting horn antenna, connected to a signal generator, was positioned approximately 15~m from the sample and oriented to illuminate the metasurface at normal incidence. At this distance, the incident field over the metasurface area can be roughly approximated by a plane wave, as further discussed in Methods. The received signal near the lower edge of the metasurface was then sampled using a subwavelength electromagnetic probe antenna connected to a spectrum analyzer, allowing us to map the local field enhancement in the region where a mobile device antenna would interact with the field.

We carried out near-field measurements along two linear scan regions parallel to the $y$-axis: one located in the metasurface plane and another parallel to it but displaced away from the metasurface by $0.5\lambda$ in the normal direction. Along each scan region, the received power was recorded at 12 equally spaced positions $y_{\rm sc}$ from $0$ to $2\lambda$, with a step of approximately $0.18\lambda$. Here, $y_{\rm sc}$ denotes the distance measured from the bottom edge of the metasurface, as shown in Fig.~\ref{fig:fig4}(a). The two scan regions were chosen to verify that the signal enhancement produced by the metasurface is not confined to a single point, but persists over a finite spatial region and therefore allows some tolerance in the position of the mobile antenna. For each scan region, we measured the received power both in the presence of the metasurface textile, $P_{\rm w}$, and in its absence, $P_{\rm w/o}$. Since the propagation  environment and the experimental configuration were kept unchanged between these two measurements, the ratio $P_{\rm w}/P_{\rm w/o}$ provides a direct measure of the received power enhancement produced by the metasurface textile.

Figure~\ref{fig:fig4}(b) shows the measured power-enhancement ratio for the fabricated textile metasurface as a function of the probe position $y_{\rm sc}$ for both scan regions. The enhancement reaches a maximum value of approximately eight, corresponding to about 9~dB, when the probe is located close to the metasurface edge. The enhancement remains robust, with values around six (approximately 8~dB) for both scan regions when the probe is within $y_{\rm sc}\leq 1.2\lambda$ from the metasurface edge. The response then gradually decreases toward $y_{\rm sc}=2\lambda$. This behavior indicates that the textile metasurface enhances the received signal over a finite spatial region rather than producing only a strongly localized scattering maximum, in agreement with the power-flow-based FOM defined in Eq.~\eqref{eq:fomp}.

When comparing the maximum measured enhancement with the simulated response of a textile metasurface of the same length ($L_y=40.4\lambda$), we observe a moderate reduction: an eightfold enhancement in the experiment compared with a simulated enhancement of 13~times. This difference can be attributed to several practical imperfections in the fabricated sample and measurement configuration, including waviness of the textile surface, quasi-planar wavefront of the incident wave, finite resolution and alignment tolerances of the screen-printing process, uncertainty in the dielectric permittivity and thickness of the textile substrate, and possible deviations from the idealized sample geometry used in simulations. In addition, the finite-size  probe antenna could perturb the local total field near the metasurface and spatially average the measured signal, reducing the field amplitude at the local maxima points. Nevertheless, the measured results clearly demonstrate a large  enhancement of received power  despite these fabrication and measurement imperfections, confirming the feasibility of the proposed textile metasurface under realistic experimental conditions.

To verify that the measured enhancement corresponds to directional routing of power by the metasurface, rather than to arbitrary local hot spots near the sample, we repeated the same measurement with the scan region positioned near the upper edge of the metasurface, as shown in Fig.~\ref{fig:fig4}(c). In this case, no noticeable enhancement is observed. This result confirms that the incident power converted into guided surface waves is predominantly directed toward the intended bottom-edge receiving region, instead of being launched symmetrically or scattered randomly around the metasurface. As a further control experiment, we performed the same measurement using a simple copper plate of the same size as the textile metasurface. Figure~\ref{fig:fig4}(d) shows that the corresponding received-power ratio remains close to unity. Therefore, the enhancement observed in Fig.~\ref{fig:fig4}(b) cannot be explained by the mere presence of a conducting surface or by edge scattering from a metallic pattern, but rather it arises from the engineered wave-conversion and power-routing functionality of the metasurface.

Finally, we performed the same set of measurements for a metasurface sample fabricated on a rigid printed circuit board (PCB) instead of a flexible textile. This control sample allows us to evaluate the same wave-conversion principle under more ideal and reproducible fabrication conditions, including a precisely defined substrate thickness and dielectric constant, a flat and mechanically stable surface, and lower material losses due to the use of a low-loss PCB substrate and a copper conductive pattern instead of printed silver ink. The geometric parameters of the PCB-based metasurface, together with the optimized values of $X_{{\rm L},n}$ and the corresponding capacitor gap dimensions $a_n$, are provided in Supplementary Section~6. Figure~\ref{fig:fig4}(e) shows the measured enhancement for this PCB metasurface, demonstrating a received-power enhancement with the maximum value of approximately 12 times the reference, corresponding to 10.8~dB. Compared with the data for textile metasurface given in Fig.~\ref{fig:fig4}(b), the enhancement peak is larger but decreases more rapidly as the probe is moved away from the metasurface edge. This difference is consistent with the more controlled and lower-loss nature of the PCB implementation, which supports a stronger and more spatially concentrated guided field near the receiving edge. By contrast, in the textile implementation, surface waviness, fabrication tolerances, dielectric permittivity uncertainty and conductor loss can broaden the enhanced-field region while reducing the peak value. The simulated response of the PCB-based metasurface predicts a power enhancement of 12.5~times, showing better agreement with the experimental result than in the textile case. This improved agreement further confirms that the residual discrepancy observed for the textile metasurface originates mainly from practical fabrication and material uncertainties rather than from the underlying metasurface design principle.

\section*{Discussion}

We have proposed and demonstrated a passive screen-printed metasurface textile for enhancing wireless links with compact user equipment. The metasurface exploits the otherwise unused surface area of clothing as an auxiliary electromagnetically functional area: it converts part of the incident field into guided surface waves and routes them toward the region where a mobile antenna can be located. In the proof-of-concept experiment at 26~GHz, the textile implementation produced an approximately eightfold received-power enhancement, while the PCB implementation reached an enhancement of approximately 12 times under more controlled fabrication conditions.


The concept is naturally user-oriented and compatible with low-cost consumer products. Since the conductive pattern is produced by screen printing, the fabrication can be scaled to large textile areas using processes similar to those already used for ordinary printed garments, with conductive ink replacing conventional graphic ink. This opens opportunities for smart clothing, detachable textile overlays, safety vests and uniforms for weak-signal environments, emergency services, industrial settings, rural connectivity, and other mission-critical communication scenarios. The same principle could also be applied to non-wearable surfaces, such as tables, walls or rigid panels, where incident power collected over a large area could be passively channeled toward a prescribed receiver location.

The present prototype demonstrates the principle for a single frequency, TE polarization, normal incidence, and concentration of power into an extended linear receiving region. Future work can extend the same design framework to dual-polarized operation and fully two-dimensional routing on realistic garments. Moreover, focusing the launched surface waves to a single point, rather than to a line, could further increase the local field and provide stronger signal enhancement for compact wireless devices.


\section*{Methods}



\paragraph{Printing of metasurfaces on textile.}
The smart textile was fabricated through a sequential printing and lamination process. A $100~\mu$m-thick TPU film, LPT5302 MR7X by Covestro, was used as an intermediate layer for holding the conductive metasurface pattern. This film provided a mechanically compliant and thermally processable support layer, allowing the printed structure to be subsequently integrated with the textile. The conductive pattern was formed using a silver flake-based ink, Metalon HPS U57B from NovaCentrix, which was deposited onto the TPU film by automatic screen printing using an Ekra X5 Professional printer. After printing, the samples were annealed at $140~^\circ$C for 30~min to remove residual solvent, promote consolidation of the silver flakes, and reduce the electrical resistance of the printed traces. The TPU films with imprinted metasurface pattern were then transferred to the textile using a heated bed laminator, FZLCB5 2. Lamination was performed at $170~^\circ$C for 15 min under a constant pressure of 45 psi, equivalent to 0.31 MPa. During this step, the TPU softened and partially penetrated into the textile matrix, forming a mechanically robust bond between the printed metasurface layer and the fabric. This process produced a flexible textile integrated metasurface while preserving the printed conductive pattern and ensuring sufficient adhesion for wearable operation.

\paragraph{Ink resistivity characterization.}
To select a conductive ink, several commercial silver inks were compared in terms of bulk resistivity $\rho$ and printing accuracy. A test pattern consisting of line arrays with multiple widths and spacings was designed. All screen-printed samples were fabricated using the same printing process, and drying or curing was performed following each vendor's datasheet. Three inks were evaluated: Metalon HPS-U57B by Novacentrix, CI-1036 by Engineered Materials Systems, and Saral StretchSilver 500 by Saralon. For reference, a subset of samples was also prepared by the bar coating (alternative to screen printing) method.
All test samples were printed using an Ekra SERIO~8000 large-format stencil printer. The screen was mounted on a 73~cm $\times$ 73~cm aluminum frame with a 40~mm $\times$ 40~mm profile and tensioned at 22.5$^\circ$. NBC UX~79--045 polyester mesh with roll width 136~cm was used, and the stencil was exposed from A4 positive film. The same screen and printer settings were used for all ink candidates.
The screen printing method produced more accurate features than the bar coating method. The sheet resistance $R_{\rm s}$ was measured using an equally spaced four-probe method, and the film thickness $t$ was measured  optically. The bulk resistivity was then calculated as $\rho=R_{\rm s} t$.
Because the final garment stack uses TPU lamination to bond the printed conductor to the fabric, $R_{\rm s}$ was measured before and after lamination. Among the evaluated inks, Metalon exhibited the lowest bulk resistivity, with a measured value of $\rho = 0.0716~\Omega\cdot\mu\mathrm{m}$. In comparison, CI-1036 showed a higher resistivity of $\rho = 0.781~\Omega\cdot\mu\mathrm{m}$, while Saral~500 exhibited the highest resistivity, $\rho = 13.49~\Omega\cdot\mu\mathrm{m}$.
In terms of printing accuracy, the Metalon and CI-1036 inks achieved minimum practical linewidths of approximately 170~$\mu$m and 150~$\mu$m, respectively. In contrast, Saral~500 ink was limited to a minimum practical linewidth of approximately 0.5~mm, indicating lower printing resolution and making it less suitable for printing fine conductive features. In addition, Saral~500 showed poor short-term process stability: after approximately 1~min of exposure to ambient air before printing, the ink visibly degraded, which further reduced its suitability for the screen-printing process used in this work.
Based on these results, the Metalon ink was selected for the sample fabrication. 

\paragraph{Experimental characterization.}
The electromagnetic response of the metasurface textile was experimentally characterized at 26~GHz. A continuous-wave signal generated by a Rohde \& Schwarz SMR~60 signal generator was fed to a rectangular horn antenna with $15$~dBi gain, which illuminated the sample under test at normal incidence.  The sample was mounted on a low-reflection stand at a distance of 16~m from the transmitting antenna. It should be noted that this distance was still shorter than the Fraunhofer distance, $2(L_x^2+L_y^2)/\lambda$, of the tested metasurface 
sample, but it was the largest separation achievable in our experimental setup. Therefore, the incident field is not an ideal plane wave over the entire sample area, but its variation across the sample remains sufficiently gradual for the field to be treated as approximately plane-wave illumination in the present experiment.
The local field near the receiving edge of the sample was measured using a subwavelength electric-field probe formed from a coaxial cable with the outer conductor and dielectric removed over a length of 2~mm. The probe output was connected to a low-noise amplifier, and the amplified signal was recorded using an Agilent 8564EC spectrum analyzer. Reference measurements were performed under identical environmental and instrumental conditions after removing the metasurface sample from the setup, allowing the measured signal to be normalized to the measurement of the corresponding free-space case.



\section*{Code availability}
Code supporting the findings of this study is available from the corresponding author upon request.

\section*{Acknowledgements}
We thank Dr. Pouyan Rezapoor for his assistance and valuable insights during the experimental stage.


\section*{Funding}
V. A. discloses support for the research of this work
from the Research Council of Finland [Projects No. 365679 and 371367],
the Finnish Foundation for Technology Promotion, the Research Council
of Finland Flagship Program Photonics Research and Innovation (PREIN)
[Decision No. 346529], and acknowledges the use of MIDAS infrastructure
of Aalto School of Electrical Engineering. X. W. acknowledges financial support from the Fundamental Research Funds for the Central Universities (Project No. 3072026RL2501) and the National Natural Science Foundation of China (Grant No. 62541116).
The remaining
authors declare no relevant funding.


\section*{Competing interests}
The authors declare no competing interests.

\bibliographystyle{unsrtnat}
\bibliography{References}

\end{document}